\documentclass[twocolumn,showpacs,prb,amsfonts,amsmath,amssymb]{revtex4}
\usepackage{bbm}
\usepackage{mathrsfs}
\usepackage{graphicx}
\usepackage{dcolumn}
\usepackage{array}
\usepackage{bm}
\begin{document}
\title{Spin states and persistent currents in mesoscopic rings:\\spin-orbit interactions}
\author{J. S. Sheng}
\author{Kai Chang}
\email[Author to whom correspondence should be addressed. Electronic
address: ]{kchang@red.semi.ac.cn}
\affiliation{SKLSM, Institute of
Semiconductors, Chinese Academy of Sciences, P. O. Box 912, Beijing
100083, China}

\begin{abstract}
We investigate theoretically electron spin states in one dimensional
(1D) and two dimensional (2D) hard-wall mesoscopic rings in the
presence of both the Rashba spin-orbit interaction (RSOI) and the
Dresselhaus spin-orbit interaction (DSOI) in a perpendicular
magnetic field. The Hamiltonian of the RSOI alone is mathematically
equivalent to that of the DSOI alone using an SU(2) spin rotation
transformation. Our theoretical results show that the interplay
between the RSOI and DSOI results in an effective periodic
potential, which consequently leads to gaps in the energy spectrum.
This periodic potential also weakens and smoothens the oscillations
of the persistent charge current (CC) and spin current (SC) and
results in the localization of electrons. For a 2D ring with a
finite width, higher radial modes destroy the periodic oscillations
of persistent currents.

\end{abstract}

 \maketitle

\section{Introduction}

In recent years, the spin-orbit interaction (SOI) in low-dimensional
semiconductor structures has attracted considerable attention due to
its potential applications in spintronic
devices.~\cite{Wolf-2001-1488,Tsitsishvili-2004-115316} There are
two types of SOI in conventional semiconductors. One is the
Dresselhaus spin-orbit interaction (DSOI) induced by bulk inversion
asymmetry,~\cite{Dresselhaus-1955-580} and the other is the Rashba
spin-orbit interaction (RSOI) induced by structure inversion
asymmetry.~\cite{Rashba-1960-1109,Bychkov-1984-6039} The strength of
the RSOI can be tuned by external gate voltages or asymmetric
doping. Recently, the intrinsic spin Hall effect (SHE) in a
spin-orbit coupled three-dimensional p-doped
semiconductor~\cite{Murakami-2003-1348} and in a Rashba spin-orbit
coupled two-dimensional electron system~\cite{Sinova-2004-126603}
was predicted theoretically. It provides us a possibility to
generate the spin current (SC) electrically without the use of
ferromagnetic metal or a magnetic field.

Recently advanced growth techniques make it possible to fabricate
high quality semiconductor rings.~\cite{Fuhrer-2001-822} A quantum
ring exhibits the intriguing spin interference phenomenon because of
its unique topology. The persistent CC in mesoscopic rings threaded
by a magnetic U(1) flux has been studied extensively, neglecting the
spin degree of freedom of the
electron.~\cite{Buttiker-1983-365,Wendler-1994-4642,Chakraborty-1994-8460}
It has been experimentally observed both in a gold
ring~\cite{Chandrasekhar-1991-3578} and in a GaAs-AlGaAs
ring~\cite{Mailly-1993-2020} using standard SQUID magnetometry. As
for the persistent SC, the SC-induced electric field that was
predicted by several
authors\cite{Meier-2003-167204,Schutz-2003-017205} may contribute to
the successful measurement of the persistent SC in mesoscopic rings
in future. The electronic structures and magneto-transport
properties of 1D rings with the RSOI alone have attracted
considerable
interest.~\cite{Splettstoesser-2003-165341,Moln'ar-2004-155335,Souma-2005-106602}
Since the strength of the DSOI in thin quantum wells is comparable
with that of the RSOI,~\cite{Lommer-1988-728} one should consider
both of the SOI's in low dimensional structures. A few works have
been done on the effects of the competition between these two types
of SOI on the transport properties of
2DEG.~\cite{Ganichev-2004-256601,
Chang-2005-085315,Yang-2006-045303} The effects of the interplay
between the RSOI and DSOI on the spin states and persistent currents
(CC and SC) in mesoscopic rings are highly desirable.

In this paper, we investigate theoretically the spin states and
persistent CC and SC in mesoscopic rings under a uniform
perpendicular magnetic field in the presence of both RSOI and DSOI.
We find that the persistent CC and SC, charge density distribution,
and local spin orientation are very sensitive to the strength of the
RSOI and DSOI. The interplay between the RSOI and DSOI leads to an
effective periodic potential. This potential has significant effects
on the physical properties of mesoscopic rings, e.g., the energy
gaps, the localization of electrons, and weakening and smoothening
of persistent CC and SC. Five different cases are considered: (1) a
1D ring with RSOI alone; (2) a 1D ring with DSOI alone; (3) a 1D
ring with RSOI and DSOI of equal strengths; (4) a 1D ring with RSOI
and DSOI of different strengths; (5) finite-width effects on the
energy spectrum, charge density distribution, the persistent CC and
SC. The eigenstates of cases one and two are analytically solved and
can be connected by a unitary transformation. The paper is organized
as follows. In Sec.~\ref{sec:theory} the theoretical model is
presented. The numerical results and discussions are given in
Sec.~\ref{sec:results}. Finally, we give a brief conclusion in
Sec.~\ref{sec:summary}.

\section{\label{sec:theory}THEORETICAL MODEL}

\subsection{Hamiltonian}

\begin{figure}[ptb]
\includegraphics[width=\columnwidth]{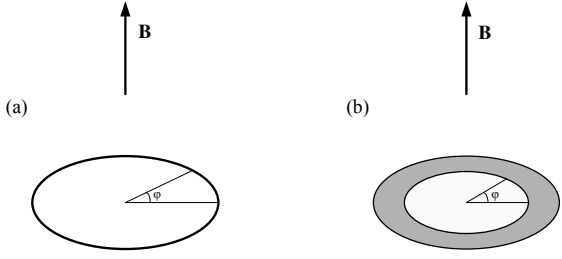}\caption{Schematic diagrams for
1D ideal ring (a) and 2D hard-wall ring (b).} \label{fig:sche}
\end{figure}

In the presence of both RSOI and DSOI, the single-particle Hamiltonian for an
electron in a finite-width ring (see Fig.~\ref{fig:sche}(b)) under a uniform
perpendicular magnetic field reads
\begin{align}
H  &  =\frac{\hbar^{2}k^{2}}{2m^{\ast}}+\alpha(\sigma_{x}k_{y}-\sigma_{y}
k_{x})+\beta(\sigma_{x}k_{x}-\sigma_{y}k_{y})\nonumber \\
&  +\frac{1}{2}g^{\ast}\mu_{B}B\sigma_{z}+V(r),\label{hami}%
\end{align}
where $\bm{k}=-i\bm{\triangledown}+e\bm{A}/\hbar$. $\bm{A}
(\bm{r})=B/2(-y,x,0)$ is the vector potential. $m^{\ast}$ is the
electron effective mass. The fourth term describes the Zeeman
splitting with Bohr magneton $\mu_{B}=e\hbar/2m_{e}$ and the
effective $g$ factor $g^{\ast}$. $\sigma_{i}(i=x,y,z)$ are the Pauli
matrices. $\alpha$ and $\beta$ specify the RSOI and DSOI strengths,
respectively. $V(r)$ is the radial confining potential,
\begin{equation}
\label{potential}V(r) =\left \{
\begin{array}
[c]{cc}%
0 & r_{1} \leqslant r \leqslant r_{2}\\
\infty & \text{otherwise}%
\end{array}
\right. ,
\end{equation}
where $r_{1}$ and $r_{2}$ are the inner and outer radii of the ring, respectively.

In the following we take the average radius $a=(r_{1}+r_{2})/2$ as
the length unit and $E_{0}=\hbar^{2}/2m^{\ast}a^{2}$ as the energy
unit. The dimensionless Hamiltonian in the polar coordinates becomes
\begin{equation}
\label{mhami}H=\begin{bmatrix} H_{k}+V(r)+\bar{g}b/2 & \bar{\beta}k_{+}+i\bar{\alpha}k_{-}\\ \bar{\beta}k_{-}-i\bar{\alpha}k_{+} & H_{k}+V(r)-\bar{g}b/2 \end{bmatrix},
\end{equation}
where $H_{k}=(\bm{e}_{r}k_{r}+\bm{e}_{\varphi}k_{\varphi})^{2}$ is
the dimensionless kinetic term; $k_{\pm}=k_{x}\pm ik_{y}=e^{\pm
i\varphi}(k_{r}\pm ik_{\varphi})$, with
$k_{r}=-i\frac{\partial}{\partial r}$ and $k_{\varphi} =-\frac{i}
{r}\frac{\partial}{\partial \varphi}+\frac{b}{4}r$; $b=\hbar
eB/m^{\ast}E_{0}$ is the dimensionless magnetic field; $\bar{\alpha}%
(\bar{\beta})=\alpha(\beta)/E_{0}a$ specifies the dimensionless RSOI
(DSOI) strength; and $\bar{g}=g^{\ast}m^{\ast}/2m_{e}$ is the
dimensionless $g$ factor.

The wavefunction $\Psi(\bm{r})$ of an electron in the ring can be
expanded as
\begin{equation}
\Psi(\bm{r}) =\sum_{nm\sigma}a_{nm\sigma}R_{n}(r) \Theta_{m}( \varphi
)\chi_{\sigma}( s_{z}),\label{expansion}%
\end{equation}
\begin{align}
\lefteqn{R_{n}(r)\Theta_{m}(\varphi)\chi_{\sigma}(s_{z})= {}}\nonumber \\
&   {} \sqrt{\frac{2}{dr}}\sin \left[ \frac{n\pi}{d}(r-r_{1})\right]
\cdot \frac{1}{\sqrt{2\pi}}e^{im\varphi}\cdot \chi_{\sigma}( s_{z}%
),\label{basicset}%
\end{align}
where $d=r_{2}-r_{1}$ is the width of the ring and $\chi_{\sigma}%
(s_{z})(\sigma=\pm1)$ are the eigenvectors of $s_{z}$.

Most previous theoretical studies on mesoscopic rings are based on
the Hamiltonian of a 1D ring (see Fig.~\ref{fig:sche}(a)), which can
be obtained by simply disregarding all the terms proportional to
derivatives with respect to $r$ in the 2D Hamiltonian. But this
conventional procedure leads to a non-hermitian Hamiltonian in the
presence of RSOI or DSOI.~\cite{Meijer-2002-033107} We can obtain a
hermitian Hamiltonian including both RSOI and DSOI following
Ref.~\onlinecite{Meijer-2002-033107}. The stationary Schr\"{o}dinger
equation is
\begin{equation}
H\Psi=E\Psi.\label{schr}%
\end{equation}
The dimensionless 1D Hamiltonian including both RSOI and DSOI reads
\begin{align}
H  & =\left[ -i\frac{\partial}{\partial \varphi}+\frac{\Phi}{\Phi_{0}}%
+\frac{\bar{\alpha}}{2}\sigma_{r}-\frac{\bar{\beta}}{2}\sigma_{\varphi}(
-\varphi)\right] ^{2}\nonumber \\
& -\frac{\bar{\alpha}^{2}+\bar{\beta}^{2}}{4}+\frac{\bar{\alpha}\bar{\beta}
}{2}\sin2\varphi+\frac{1}{2}\bar{g}b\sigma_{z},\label{1dhami}%
\end{align}
where $\sigma_{r}=\cos \varphi \sigma_{x}+\sin \varphi \sigma_{y}$,
$\sigma _{\varphi}=\cos \varphi \sigma_{y}-\sin \varphi \sigma_{x}$,
$\Phi=B\pi a^{2}$ is the magnetic flux threading the ring, and
$\Phi_{0}=h/e$ is the flux unit. Notice that there is a periodic
potential term ($\frac {\bar{\alpha}\bar{\beta} }{2}\sin2\varphi$)
induced by the interplay between the RSOI and DSOI.

We introduce a vector function $S(\bm{r})$ to describe the local
spin orientation of a specific eigenstate $\Psi$ in a 1D ring:
\begin{equation}
S(\bm{r})=\Psi^{\dag}\bm{s}\Psi=\Psi^{\dag}s_{x}\Psi \mathbf{e}_{x}+\Psi^{\dag
}s_{y} \Psi \mathbf{e}_{y}+\Psi^{\dag}s_{z}\Psi \mathbf{e}_{z} .\label{lso}%
\end{equation}

When the coupling strength $\bar{\alpha}$ or $\bar{\beta}$ vanishes, the
$\sin2\varphi$ potential accordingly disappears and the analytical solution to
the eigenvalue problem is available (see Appendix for details). Generally we
have to solve Eq.~(\ref{schr}) numerically when $\bar \alpha \neq0$ and
$\bar \beta \neq0$.

\subsection{Persistent currents}

The charge density operator and the charge current density operator are
\begin{equation}
\label{coperators}%
\begin{split}
\hat{\rho}(\bm{r}^{\prime}) & =-e\delta( \bm{r}^{\prime}-\bm{r})\\
\hat{\bm{j}}_{c}(\bm{r}^{\prime})  & = \frac{1}{2}[ \hat{\rho}( \bm{r}^{\prime
})\hat{\bm{v}}+\hat{\bm{v}}\hat{\rho}( \bm{r}^{\prime}))],
\end{split}
\end{equation}
where $\bm{r}^{\prime}$ refers to the field coordinates and $\bm{r}$
the coordinates of the electron. We can also introduce the spin
density operator and spin current density
operator~\cite{Splettstoesser-2003-165341} as
\begin{equation}
\label{soperators}%
\begin{split}
\hat{\bm{S}}(\bm{r}^{\prime}) & =\frac{\hbar}{2} \hat{\bm{\sigma}}%
\delta(\bm{r}^{\prime}-\bm{r})\\
\hat{\bm{j}}_{s}(\bm{r}^{\prime}) & =\frac{1}{2} [\hat{\bm{S}}(\bm{r}^{\prime
}) \hat{\bm{v}}+\hat{\bm{v}}\hat{\bm{S}}(\bm{r}^{\prime})],
\end{split}
\end{equation}
where $\hat{\bm{\sigma}}=\hat{\sigma}_{x}\mathbf{e}_{x}+\hat{\sigma}%
_{y}\mathbf{e}_{y}+\hat{\sigma}_{z}\mathbf{e}_{z}$ is the vector of
the Pauli matrices. The charge current density and spin current
density can be obtained by calculating the expectation values of the
corresponding operators:
\begin{equation}
\label{cdexp}
\begin{split}
\bm{j}_{c}(\bm{r}^{\prime})  &  =\left \langle \Psi \right \vert \hat{\bm{j}}_{c}
\left \vert \Psi \right \rangle = -e\operatorname{Re}\left \{  \Psi^{\dag}(
\bm{r}^{\prime}) \hat{\bm{v}}^{\prime}\Psi(\bm{r}^{\prime}) \right \} \\
\bm{j}_{s}(\bm{r}^{\prime})  &  =\left \langle \Psi \right \vert \hat{\bm{j}}_{s}
\left \vert \Psi \right \rangle = \operatorname{Re}\left \{  \Psi^{\dag}(
\bm{r}^{\prime}) \hat{\bm{v}}^{\prime}\hat{\bm{s}}\Psi( \bm{r}^{\prime
})\right \} ,
\end{split}
\end{equation}
where $\Psi(\mathbf{r})$ is the wavefunction of an electron in the
ring. For convenience, we note
$\bm{r}^{\prime},\hat{\bm{v}}^{\prime}$ as $\bm{r},\hat {\bm{v}}$
hereafter.

The $\varphi$-component of the velocity operator associated with the
Hamiltonian in Eq.~(\ref{hami}) is
\begin{equation}
\hat{\bm{v}}_{\varphi}=\mathbf{e}_{\varphi}\left[  \frac{\hbar} {im^{\ast}%
r}\frac{\partial}{\partial \varphi}+\frac{eBr}{2m^{\ast} }+\frac{\alpha}{\hbar
}\sigma_{r}-\frac{\beta}{\hbar}\sigma_{\varphi}( -\varphi) \right]
.\label{velocity}%
\end{equation}
The azimuthal (spin or charge) current can be defined
as~\cite{Wendler-1994-4642}
\begin{equation}
I_{\varphi}=\frac{1}{2\pi}\int_{0}^{2\pi} \! \! \! \mathrm{d} \varphi \int
_{r_{1}}^{r_{2}} \! \! \! \mathrm{d} r j_{\varphi}( \bm{r}).\label{current}%
\end{equation}

We ignore the Coulomb interaction between electrons in this work. At
the low temperature, $N$ electrons will occupy the lowest $N$ levels
of the energy spectrum. The total (charge or spin) current is the
summation over all occupied
levels.~\cite{Splettstoesser-2003-165341}

For a 1D ring, the eigenstates could be expanded in the basis set
$\Theta _{m}(\varphi) =\exp( im\varphi)/\sqrt{2\pi}$ which is much
simpler than that in Eq.~(\ref{expansion}), and we can get the
azimuthal component of the velocity operator in a 1D ring by
specifying the variable $r$ as the constant $a$ in
Eq.~(\ref{velocity}).

Most of the previous investigations of the persistent CC in
mesoscopic rings are based on the well-known formula
$I_{n}=-\partial E_{n}/\partial \Phi$, where $I_{n}$ denotes the
contribution to the persistent CC from the $n$th state and $\Phi$ is
the magnetic flux through the
ring.~\cite{Splettstoesser-2003-165341,Chakraborty-NQR:ANP-2003,Lozano-2005-205315}
In this paper we calculate the persistent CC and SC via
Eq.~(\ref{cdexp}). Note that for a 1D ring our results are identical
with those obtained from the formula $I_{n}=-\partial E_{n}/\partial
\Phi$.

\section{\label{sec:results}RESULTS AND DISCUSSION}

\begin{figure}[ptb]
\includegraphics[width=\columnwidth]{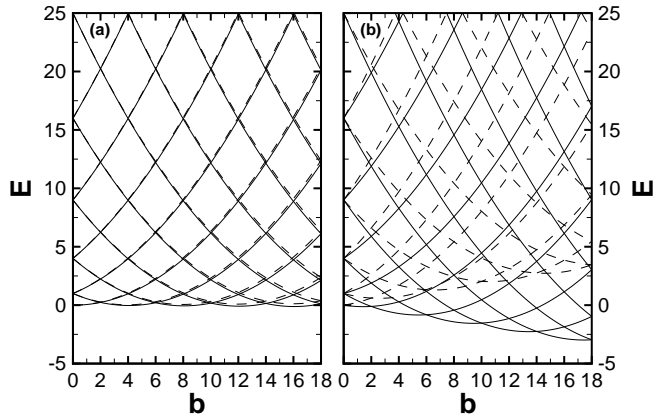}\caption{(a) The energy spectrum for 1D
GaAs rings. $\bar{\alpha }=\bar{\beta}=0$, $\bar{g}=-0.01474$; (b)
 The energy spectrum
for 1D InSb ring. $\bar{\alpha}=\bar{\beta}=0$, $\bar{g}=-0.357$. In
Fig.~\ref{fig:GaAsInsb}(a) and Fig.~\ref{fig:GaAsInsb}(b) the solid
(dashed) lines denote spin-up (spin-down)
levels.}\label{fig:GaAsInsb}
\end{figure}

We show the energy spectrum of a 1D ring in Fig.~\ref{fig:GaAsInsb}
for different $g$ factors. The relevant parameters of the materials
used in our calculation are listed in Table~\ref{tab:par}. Without
the spin-orbit couplings, the $g$ factor accounts for the spin
splitting. For a material with large $g$ factor such as InSb,
although the parabola behavior of the energy levels as functions of
the magnetic fields is still retained, the periodicity of the energy
spectrum is severely broken by the $g$ factor, especially at large
magnetic fields. For a material with a small $g$ factor, e.g. GaAs,
the Zeeman splitting is quite small even in a rather large magnetic
field.

\begin{table}[ptb]
\caption{Parameters used in our calculation are from
Ref.~\onlinecite{LB-17a-1982}.}%
\label{tab:par}%
\begin{ruledtabular}
\begin{tabular}{ccc}
& $m^{*}(m_{e})$ & $g^{*}$ \\
\hline GaAs & 0.067 & -0.44 \\
InSb & 0.014  & -51  \\
\end{tabular}
\end{ruledtabular}
\end{table}

\subsection{1D ring with RSOI alone}

As shown in the Appendix, the electron states in a 1D ring with RSOI
alone under a uniform magnetic field including the Zeeman splitting
can be solved analytically.

\begin{figure}[ptb]
\includegraphics[width=\columnwidth]{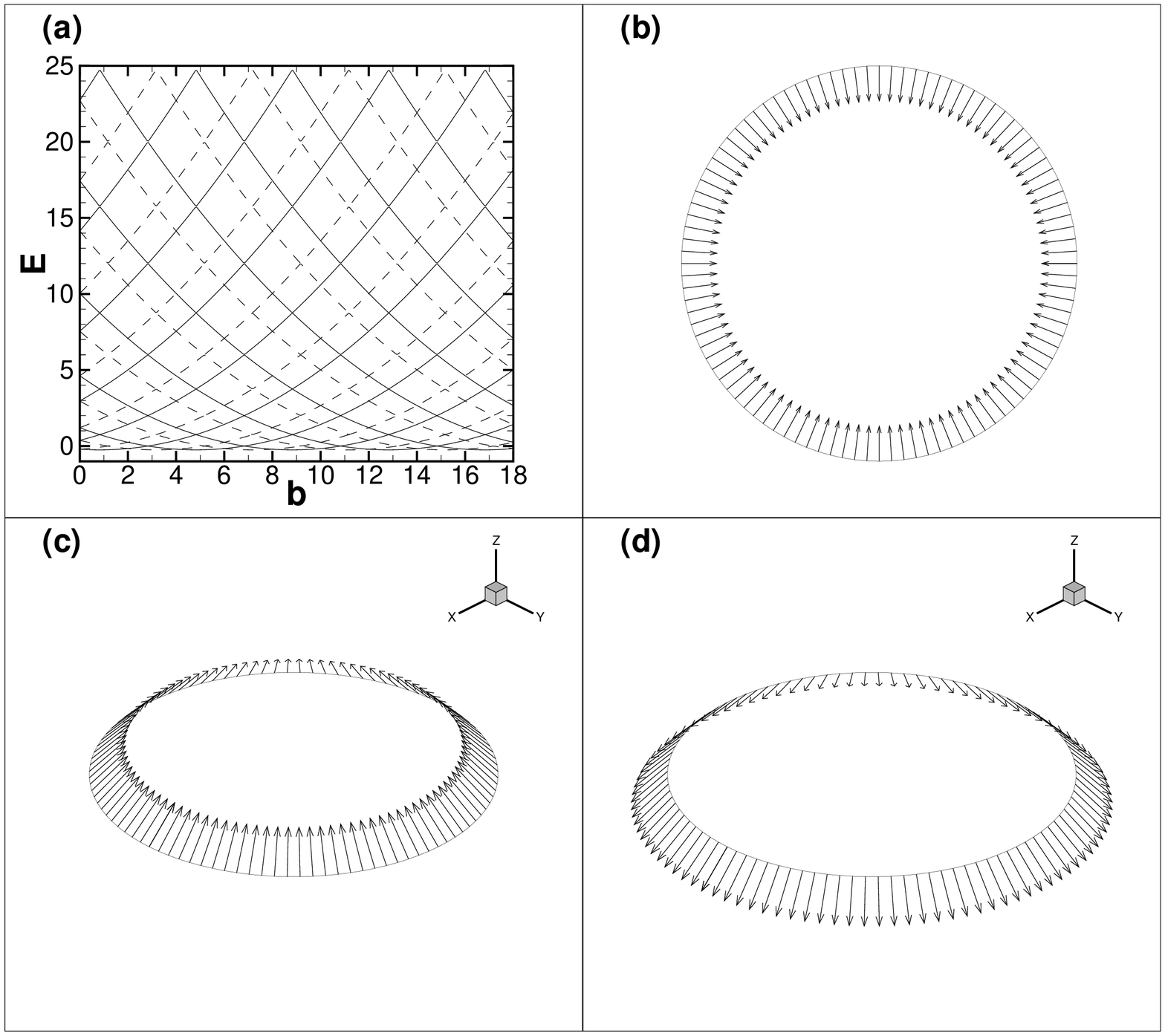}\caption{(a) Energy
spectrum of 1D ring in the presence of RSOI alone, where the solid
lines (dashed lines) denote the spin-up (spin-down) levels; (b) The
projection of $S(\bm{r})_{\uparrow}^{R}$ onto the $x$-$y$ plane; (c)
Local spin orientation for all the spin-up levels
$S(\bm{r})_{\uparrow}^{R}$; (d) Local spin orientation for all the
spin-down levels $S(\bm{r})_{\downarrow}^{R}$. $\bar{\alpha}=1$,
$\bar{\beta}=0$ and $\bar{g}=0$.}
\label{fig:alpha1beta0}%
\end{figure}

The energy spectrum in the presence of RSOI alone is plotted in
Fig.~\ref{fig:alpha1beta0}(a). When the $g$ factor is set to be
zero, the energy of each level is given by
\begin{equation}
E_{n,\sigma} =\left(
n+\frac{b}{4}+\frac{\sigma}{2}-\frac{\sigma}{2\cos \theta
}\right) ^{2}-\frac{\tan^{2}\theta}{4},\label{spectrum}%
\end{equation}
where $\theta=\arctan(\bar{\alpha})$ (Eq.~(\ref{rups}) or
Eq.~(\ref{rdowns}) represents the corresponding eigenstate so long
as $\theta_{n,\sigma}$ is replaced by $\theta$). From the above
expression we find that the spin-up and spin-down levels with the
same quantum number $n$ are separated in the $b$ axis by $4(1/\cos
\theta-1)$, and both of the spin-up and spin-down levels are pulled
down by $\tan^{2}\theta/4$ compared to the case without SOI.

For the RSOI alone case, the local spin orientation for all the
spin-up states along the ring is described as
\begin{align}
&  S(\bm{r})_{\uparrow}^{R}\nonumber \\
&  =\Psi_{n,\uparrow}^{R\dag}s_{x}\Psi_{n,\uparrow}^{R}\mathbf{e}_{x}
+\Psi_{n,\uparrow}^{R\dag}s_{y}\Psi_{n,\uparrow}^{R}\mathbf{e}_{y}
+\Psi_{n,\uparrow}^{R\dag}s_{z}\Psi_{n,\uparrow}^{R}\mathbf{e}_{z}\nonumber \\
&  =\frac{\hbar}{4\pi a}\left[  \sin( -\theta)( \cos \varphi \mathbf{e}_{x}%
+\sin \varphi \mathbf{e}_{y}) +\cos( -\theta) \mathbf{e}_{z}\right]
.\label{rlso}%
\end{align}
The local spin orientation for all the spin-down states
$S(\bm{r})_{\downarrow }^{R}$ is opposite to
$S(\bm{r})_{\uparrow}^{R}$ as shown in Fig.~\ref{fig:alpha1beta0}(c)
and Fig.~\ref{fig:alpha1beta0}(d). In this case, the oblique angle
$\theta=\arctan(\bar{\alpha})$ becomes independent from the quantum
number $n$ and the magnetic field $b$. It means that the local spin
orientations for all the spin-up (spin-down) states are the same.
When there is RSOI alone, the local spin orientation exhibits
rotational symmetry for either spin-up or spin-down states.

\begin{figure}[ptb]
\includegraphics[width=\columnwidth]{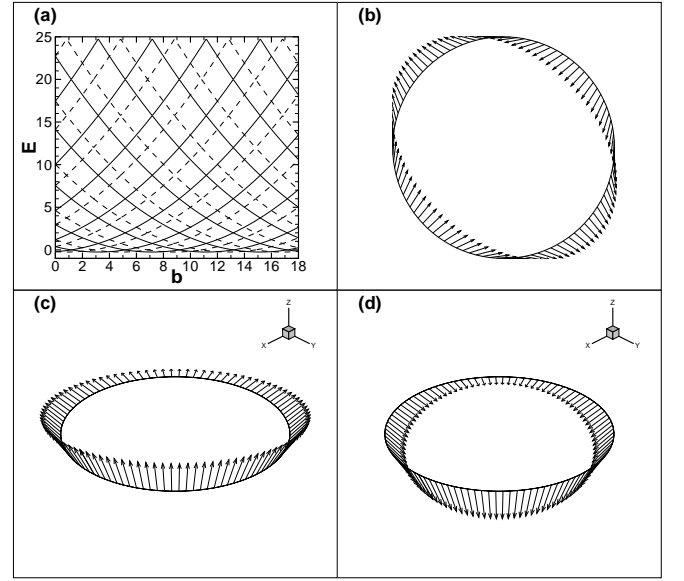}\caption{(a) Energy
spectrum of 1D ring in the presence of DSOI alone, where the solid
lines (dashed lines) denote the spin-up (spin-down) levels; (b) The
projection of $S(\bm{r})_{\uparrow}^{D}$ onto the $x$-$y$ plane; (c)
Local spin orientation for all the spin-up levels
$S(\bm{r})_{\uparrow}^{D}$; (d) Local spin orientation for all the
spin-down levels $S(\bm{r})_{\downarrow}^{D}$.
$\bar{\alpha}=0$, $\bar{\beta}=1$ and $\bar{g}=0$.}%
\label{fig:alpha0beta1}%
\end{figure}

The contributions to the persistent CC and SC from each level can be
easily obtained explicitly:
\begin{subequations}
\begin{align}
I_{n,\sigma}  &  =-\left(
n+\frac{b}{4}+\frac{\sigma}{2}-\frac{\sigma}{2\cos
\theta}\right) ,\label{cc}\\
I_{n,\sigma}^{s_{z}}  &  =\left(
n+\frac{b}{4}+\frac{\sigma}{2}-\frac{\sigma }{2\cos \theta}\right)
\sigma \cos \theta.\label{sc}
\end{align}
$I_{n,\sigma}$ ($I_{n,\sigma}^{s_{z}}$) denotes the contribution to
the persistent CC (SC) from the eigenstate $\Psi_{n,\sigma}$.
$I_{n,\sigma}$ ($I_{n,\sigma}^{_{s_{z}}}$) is in the units of
$2E_{0}/\Phi_{0}$ ($E_{0}/2\pi $). We should notice that these
expressions deduced from operators coincide with the formula
$I_{n}=-\partial E_{n}/\partial \Phi$.

\begin{figure}[ptb]
\includegraphics[width=\columnwidth]{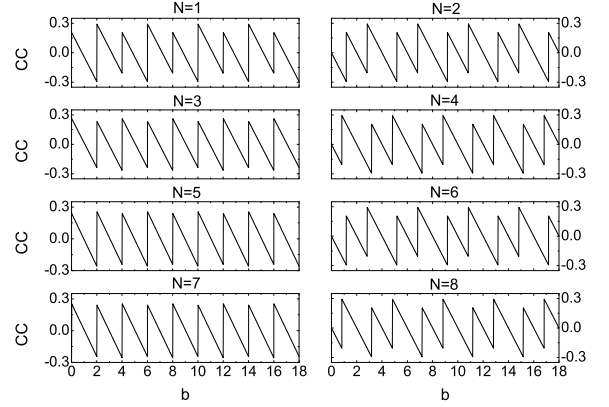}\caption{The persistent CC in a
1D ring with different numbers of electrons $N$ \textit{vs} magnetic
field $b$ while $\bar{\alpha}=1$, $\bar{\beta}=0$, $\bar{g}=0$. The
persistent CC is in
units of $2NE_{0}/\Phi_{0}$.}%
\label{fig:ccr}%
\end{figure}

Splettstoesser \textit{et al.} analyzed the persistent CC induced by
the magnetic flux in the 1D ring with both RSOI and a impurity
potential.~\cite{Splettstoesser-2003-165341} They demonstrated that
the strength of the RSOI can be extracted from the dependence of the
persistent CC on the magnetic flux. The number of electrons $N$ in
their work was assumed to be large enough. We focus on the case in
which there are few electrons in the ring (see Fig.~\ref{fig:ccr}).
We find that the persistent CC is a periodic function of $b$
exhibiting many linear segments with a slope ratio of $-1/4$ which
can be easily deduced from Eq.~(\ref{cc}). The periodicity of the
persistent CC for an arbitrary $N$ is 4, the same as that of the
energy spectrum. For an arbitrary even number of electrons
$N=2\mathbbm{n}$, the jumping amplitude is $1/2$ and the neighboring
two jumps are shifted along the CC axis. For an odd number of
electrons $N=2\mathbbm{n}+1$, there are two jumping amplitudes which
appear alternately at those points where $b=0,2,4,6,\ldots$ One
jumping amplitude is $( \mathbbm{n}-1/\cos \theta+2)/( 2\mathbbm{n}
+1)$, while the other is $( \mathbbm{n} +1/\cos \theta-1)/(
2\mathbbm{n}+1)$. When $n$ approaches infinity, the above two
amplitudes tend to $1/2$.

The dependence of the persistent SC on the magnetic field $b$ shows
interesting behavior. Turning (jumping) points in the persistent SC
oscillation for an odd (even) number of electrons are caused by the
crossing of levels with opposite (same) spins. When the magnetic
field $b$ sweeps, the persistent SC oscillation for an odd (even)
number of electrons exhibits saw-tooth (square) wave behavior and
the oscillation amplitude of the persistent SC for an even number of
electrons is bigger than that for a neighboring odd number of
electrons especially when the number of electrons increases. For an
odd number of electrons $N=2\mathbbm{n}+1$, the slope ratios of the
persistent SC are $\pm \cos \theta/4(2\mathbbm{n}+1)$ alternately.
For an even number of electrons $N=2\mathbbm{n}$, the jumping
amplitude is $\cos \theta/2$ for all jumping points.

\begin{figure}[ptb]
\includegraphics[width=\columnwidth]{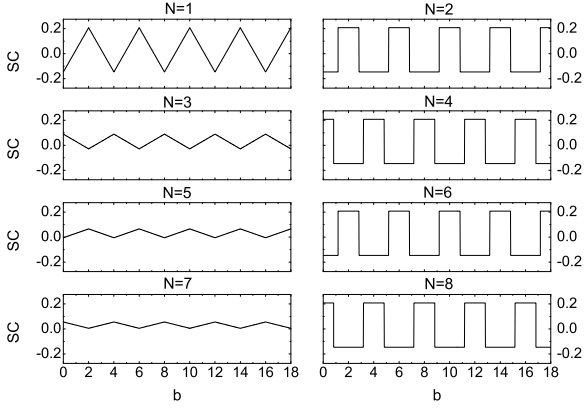}\caption{Same as
Fig.~\ref{fig:ccr}, but for the persistent SC. The persistent SC is in units
of $NE_{0}/2\pi$.}%
\label{fig:scr}%
\end{figure}

\subsection{1D ring with DSOI alone}

Now we consider a 1D ring with DSOI alone. The eigenstates of the
Hamiltonian of a 1D ring with $\bar{\alpha}=a$, $\bar{\beta}=b$, and
$\bar{g}=c$ can be connected to those with $\bar{\alpha}=b$,
$\bar{\beta}=a$, and $\bar{g}=-c$ by a unitary operator $T$ (see the
Appendix). As we will show in the next subsection, $T$ represents a
rotational transformation in spin space. $b$ and $c$ are set to be
zero and the relationship between the RSOI alone case and DSOI alone
case is specified as
\end{subequations}
\begin{subequations}
\begin{align}
E_{n,\sigma}^{D}  &  =E_{n,-\sigma}^{R},\label{connectiona}\\
\Psi_{n,\uparrow}^{D}  &  =\exp[i\pi/4]T\Psi_{n,\downarrow}^{R}%
,\label{connectionb}\\
\Psi_{n,\downarrow}^{D}  &  =\exp[-i\pi/4]T^{\dag}\Psi_{n,\uparrow}^{R}
.\label{connectionc}%
\end{align}

The energy spectrum while $\bar{\alpha}=0$, $\bar{\beta}=1$, and
$\bar{g}=0$ is plotted in Fig.~\ref{fig:alpha0beta1}(a), which is
exactly the same as that of a 1D ring with RSOI alone (see
Fig.~\ref{fig:alpha1beta0}(a)), but the spin orientations of the
corresponding eigenstates are different. This feature arises from
the sign reversal behavior of $\sigma_{z}$ under the unitary
transformation $T$.

Although the eigenstates for the RSOI alone case and those for the
DSOI alone case can be connected by the unitary transformation $T$,
the behaviors of $S(\bm{r})_{\uparrow \downarrow}^{R}$ and
$S(\bm{r})_{\uparrow \downarrow}^{D}$ are very different. In the
case with DSOI alone, spin-up (spin-down) states with different
angular quantum number $n$ or under different magnetic field $b$
share the same local spin orientation $S(\bm{r})_{\uparrow}^{D}$ $(
S(\bm{r})_{\downarrow}^{D}) $. The angle between
$S(\bm{r})_{\uparrow}^{D}$ ($S(\bm{r})_{\downarrow}^{D}$) and the
$z$-axis is $-\theta$ ($\pi-\theta$). The local spin orientations
$S(\bm{r})_{\uparrow \downarrow}^{D}$ can be obtained by
interchanging the $x$ and $y$ components of $S(\bm{r})_{\uparrow
\downarrow}^{R}$ (see Eqs.~(\ref{rlso}) and (\ref{dlso})).
\end{subequations}
\begin{align}
&  S(\bm{r})_{\uparrow}^{D}\nonumber \\
&  =\Psi_{n,\uparrow}^{D\dag}s_{x}\Psi_{n,\uparrow}^{D}\mathbf{e}_{x}
+\Psi_{n,\uparrow}^{D\dag}s_{y}\Psi_{n,\uparrow}^{D}\mathbf{e}_{y}
+\Psi_{n,\uparrow}^{D\dag}s_{z}\Psi_{n,\uparrow}^{D}\mathbf{e}_{z}\nonumber \\
& =\Psi_{n,\downarrow}^{R\dag}T^{\dag}s_{x}T\Psi_{n,\downarrow}^{R}\mathbf{e}
_{x}+\Psi_{n,\downarrow}^{R\dag}T^{\dag}s_{y}T\Psi_{n,\downarrow}^{R}
\mathbf{e}_{y}\nonumber \\
& +\Psi_{n,\downarrow}^{R\dag}T^{\dag}s_{z}T\Psi_{n,\downarrow}^{R}\mathbf{e}
_{z}\nonumber \\
& =-\Psi_{n,\downarrow}^{R\dag}s_{y}\Psi_{n,\downarrow}^{R}\mathbf{e}_{x}
-\Psi_{n,\downarrow}^{R\dag}s_{x}\Psi_{n,\downarrow}^{R}\mathbf{e}_{y}
-\Psi_{n,\downarrow}^{R\dag}s_{z}\Psi_{n,\downarrow}^{R}\mathbf{e}
_{z}\nonumber \\
&  =\Psi_{n,\uparrow}^{R\dag}s_{y}\Psi_{n,\uparrow}^{R}\mathbf{e}_{x}
+\Psi_{n,\uparrow}^{R\dag}s_{x}\Psi_{n,\uparrow}^{R}\mathbf{e}_{y}
+\Psi_{n,\uparrow}^{R\dag}s_{z}\Psi_{n,\uparrow}^{R}\mathbf{e}_{z}\nonumber \\
&  =\frac{\hbar}{4\pi a}\left[  \sin(-\theta)( \sin \varphi \mathbf{e}_{x}%
+\cos \varphi \mathbf{e}_{y}) +\cos( -\theta) \mathbf{e}_{z}\right]
.\label{dlso}%
\end{align}
This interesting feature comes from the behavior of $\sigma_{x(y)}$
under the unitary transformation $T$, i.e., $T\sigma_{x(y)
}T^{\dag}=T^{\dag}\sigma_{x(y) }T=-\sigma_{y(x)}$. The projections
of $S(\bm{r})_{\uparrow}^{R}$ and $S(\bm{r})_{\uparrow}^{D}$ onto
the $x$-$y$ plane are very different (see
Fig.~\ref{fig:alpha1beta0}(b) and Fig.~\ref{fig:alpha0beta1}(b)).
For the RSOI alone case, the vector always points along the radial
direction and the locus of the arrowhead is a circle (see
Fig.~\ref{fig:alpha1beta0}(b)). For the DSOI alone case, the vector
varies along the ring and the cylindrical symmetry is broken (see
Fig.~\ref{fig:alpha0beta1}(b)).

In the current case, the persistent CC oscillations exhibit the same
behavior as those of the RSOI alone case (see Fig.~\ref{fig:ccr})
because the corresponding levels are identical as functions of the
magnetic field $b$. We know that the contribution of each level to
the persistent SC is related not only to the magnetic field
dependence of the eigenenergy but also to the spin orientation of
the eigenstate. Since the spin-up and spin-down levels are
interchanged compared to those of the RSOI alone case, the
persistent SC in the current case can be obtained by changing the
sign of the persistent SC for the RSOI alone case.

\subsection{1D ring with equal strength RSOI and DSOI}

\begin{figure}[ptb]
\includegraphics[width=\columnwidth]{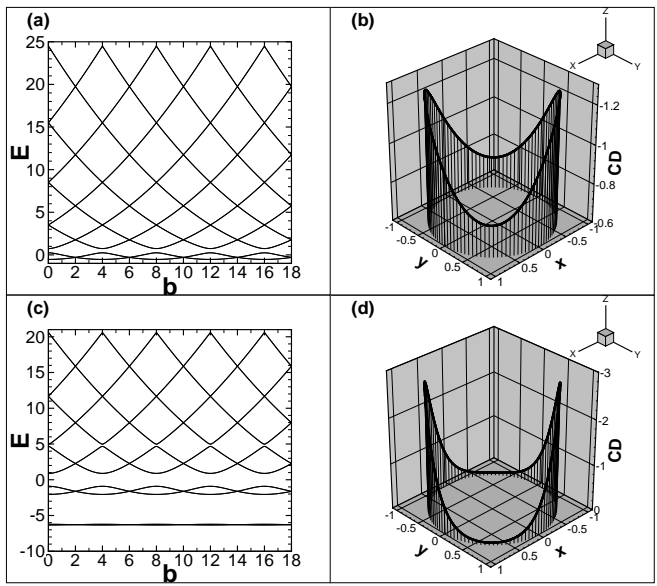}\caption{(a) Energy
spectrum of 1D ring while $\bar{\alpha}=\bar{\beta}=1$, $\bar{g}=0$; (b) The
charge density (CD) distribution of the lowest single electron state in a 1D
ring while $\bar{\alpha}=\bar{\beta}=1$, $\bar{g}=0$, $b=0$; (c) Energy
spectrum of 1D ring while $\bar{\alpha}=\bar{\beta}=3$, $\bar{g}=0$; (d) The
charge density (CD) distribution of the lowest single electron state in a 1D
ring while $\bar{\alpha} =\bar{\beta}=3$, $\bar{g}=0$, $b=0$. The charge
density is in units of $e/2\pi a$.}%
\label{fig:alphaeqbeta}%
\end{figure}

The electron energy spectra for $\bar{\alpha}=\bar{\beta}=1$,
$\bar{g}=0$ and $\bar{\alpha}=\bar{\beta}=3$, $\bar{g}=0$ are
plotted in Figs.~\ref{fig:alphaeqbeta}(a) and
\ref{fig:alphaeqbeta}(c), respectively. We can see that the energy
spectra are spin degenerate. The spin degeneracy comes from the
symmetry of the Hamiltonian when $\bar{\alpha}=\bar{\beta}$. In this
paper, we use the unitary operator $T$ (see Appendix) to describe
the symmetry of the 1D Hamiltonian. When
$\bar{\alpha}=\bar{\beta}\neq0$ and $\bar{g}=0$,
$THT^{\dag}=T^{\dag}HT=H$. If we have $H\Psi=E\Psi$, in which $\Psi$
is an eigenstate for the eigenenergy $E$, then the states $T\Psi$
and $T^{\dag}\Psi$ are also eigenstates and are equivalent to each
other. Thus the energy levels are twofold degenerate. The operator
for a $\varphi$ rotation around the unit vector $\mathbf{n}$ in the
Hilbert space reads
\begin{equation}
D(\mathbf{n},\varphi) =e^{-i\varphi \mathbf{n}\cdot \bm{L}/ \hbar}\otimes
e^{-i\varphi \mathbf{n}\cdot \bm{s}/\hbar},\label{doperator}%
\end{equation}
where $\bm{L}$ and $\bm{s}$ denote the orbital and spin angular
momentum operators, respectively. It is easy to demonstrate that the
unitary operator $T$ can be written as $\exp[-i\pi \mathbf{n}_{1}
\cdot \bm{s}/\hbar]$ with $\mathbf{n}_{1}=(
1/\sqrt{2},-1/\sqrt{2},0)$. Then $T$ is actually a rotation operator
in the spin space. For a quantum ring in the $x$-$y$ plane, the
orbital angular momentum vector $\bm{L}$ $=\bm{r}\times \bm{p}$
points along the $z$-axis, and therefore $\mathbf{n}_{1}\cdot
\bm{L}=0$. Thus the unitary operator $T$ is eventually a rotation
operator in the whole Hilbert space:
\begin{equation}
T=D(\mathbf{n}_{1},\pi).\label{toperator}%
\end{equation}
That means the unitary transformation $T$ ($T^{\dag}$) is actually a
rotation by $\pi$ ($-\pi$) around $\mathbf{n}_{1}$. Similarly, there
are also symmetric operations corresponding to $\pi$ and $-\pi$
rotations around $\mathbf{n}_{2}=( 1/\sqrt{2},1/\sqrt{2},0)$ for the
Hamiltonian with $\bar{\alpha}=-\bar{\beta}$. We need to stress that
these symmetric operations also exist for a two-dimensional electron
gas with equal strength RSOI and DSOI $( \bar{\alpha}=\pm
\bar{\beta})$ and $\bar{g}=0$.

Besides the spin degeneracy, it is interesting to find that gaps
appear in the energy spectra. In order to understand this feature,
we transform the original Hamiltonian to a simple form via a unitary
transformation $A$ as follows:
\begin{align}
A  &  =\frac{1}{\sqrt{2}}\, \, \centerdot \nonumber \\
&
\begin{bmatrix} \exp[-i\bar{\alpha}f(\varphi)] & \exp[i\bar{\alpha}f(\varphi)] \\ \exp[-i\pi/4]\exp[-i\bar{\alpha}f(\varphi)] & -\exp[-i\pi/4] \exp[i\bar{\alpha}f(\varphi)] \end{bmatrix},\label{aoperator}%
\end{align}
with $f(\varphi) =\sin(\varphi+\pi/4) $. The original electron Hamiltonian
with $\bar{\alpha}=\bar{\beta}$ and $\bar{g}=0$ reads
\begin{align}
H  &  =\left[ -i\frac{\partial}{\partial \varphi}+\frac{b}{4}+\frac{\bar
{\alpha}} {2}\sigma_{r}-\frac{\bar{\alpha}}{2}\sigma_{\varphi}\left(
-\varphi \right)  \right] ^{2}\nonumber \\
&  -\frac{\bar{\alpha}^{2}}{2}+\frac{\bar{\alpha}^{2}}{2}\sin2\varphi
.\label{dehami}%
\end{align}
After the transformation, the Hamiltonian becomes
\begin{equation}
H^{\prime}=A^{\dag}HA=\left( -i\frac{\partial}{\partial \varphi}+\frac{b}
{4}\right) ^{2}-\frac{\bar{\alpha}^{2}}{2}+\frac{\bar{\alpha}^{2}}{2}
\sin2\varphi.\label{simhami}%
\end{equation}

That means the Hamiltonian of a 1D ring with equal strength RSOI and
DSOI and zero $g$ factor is equivalent to that of a 1D ring with a
periodic potential alone (see the last term in Eq.~(\ref{simhami})).
The potential height is proportional to the square of the SOI
strength, and the average of the potential shifts down by about
$\bar{\alpha}^{2}/2$. The eigenvectors of Eq.~(\ref{simhami}) are
actually the periodic solutions of the Mathieu
equation.~\cite{Gradshteyn-ToISaP-1980} The energy gaps, which are
proportional to the potential height, decrease with decreasing SOI
strengths, especially for higher gaps (see
Fig.~\ref{fig:alphaeqbeta}(a)). When the strengths of SOI are fixed,
the higher energy gaps are narrower than the lower ones because
there is less influence from the potential.

When only one type of SOI (RSOI or DSOI) exists, the charge density
distribution will be constant along the ring. But the charge density
distribution becomes localized along the ring when both RSOI and
DSOI are taken into account. This localization arises from the
effective periodic potential, whose height is determined by the
product of the strengths of RSOI and DSOI (see Eq.~(\ref{1dhami})).
Therefore large SOI strengths lead to strong electron localization.
The absolute value of the charge density exhibits maxima at the
valleys of the $\sin2\varphi$ potential ($\varphi=3\pi/4$ or
$\varphi=7\pi/4$) and minima at the peaks of the $\sin2\varphi$
potential ($\varphi=\pi/4$ or $\varphi=5\pi/4$) (see
Fig.~\ref{fig:alphaeqbeta}(b) and Fig.~\ref{fig:alphaeqbeta}(d)).

\begin{figure}[ptb]
\includegraphics[width=\columnwidth]{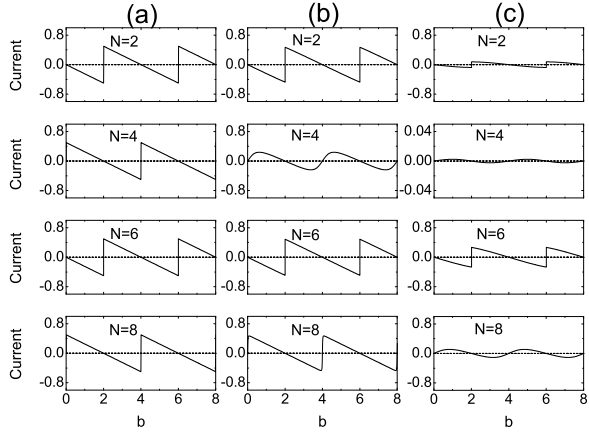}\caption{The persistent
CC (solid lines) and SC (dashed lines) with different even numbers
of electrons $N$
\textit{vs} magnetic field in the degenerate cases (a) $\bar{\alpha}%
=\bar{\beta}=\bar{g}=0$; (b) $\bar{\alpha}=\bar{\beta}=1$,
$\bar{g}=0$; (c) $\bar{\alpha}=\bar{\beta}=3$, $\bar{g}=0$. The
persistent CC (SC) is in units of $2NE_{0}/\Phi_{0}$
($NE_{0}/2\pi$).} \label{fig:currentdege}
\end{figure}

In the spin degenerate case, we cannot define the local spin
orientation $S(\bm{r})$ for an eigenenergy level and the persistent
SC for an odd number of electrons because of the uncertainty of the
eigenvectors. The persistent CC oscillation for an odd number of
electrons $N=2\mathbbm{n}+1$ is simply the arithmetic average of
that for $N=2\mathbbm{n}$ and that for $N=2\mathbbm{n}+2$ in a spin
degenerate case. Thus we compare the persistent current (CC and SC)
oscillations in the three degenerate cases:
$\bar{\alpha}=\bar{\beta}=0$, $\bar{\alpha}=\bar{\beta}=1$, and
$\bar{\alpha }=\bar{\beta}=3$ only for even numbers of electrons
(see Fig.~\ref{fig:currentdege}). The persistent SC is zero in the
two degenerate cases for even numbers of electrons. The
$\sin2\varphi$ potential in Eq.~(\ref{1dhami}) accounts for the
flatter magnetic dispersion as well as gaps in the energy spectrum
when $\bar{\alpha}=\bar{\beta}\neq0$. Since the contribution to the
persistent CC from an energy level is actually determined by the
dependence of the energy level on magnetic field, the oscillation of
the persistent CC for the case $\bar{\alpha}=\bar{\beta}\neq0$ is
smoother and smaller than that for the case
$\bar{\alpha}=\bar{\beta}=0$. The interplay between the RSOI and
DSOI smoothens and weakens the persistent CC oscillation most
obviously when the Fermi level locates near the lowest gap (see the
panels labeled $N=4$ in Fig.~\ref{fig:currentdege}(b) and
Fig.~\ref{fig:currentdege}(c)). While the number of electrons
increases, the oscillation of the persistent CC becomes sharp again
since the higher gaps become smaller. This smoothening and weakening
effect can even be found again for a large number of electrons when
the SOI strengths increase (see Fig.~\ref{fig:currentdege}(c)).

\subsection{1D ring with different strength RSOI and DSOI}

\begin{figure}[ptb]
\includegraphics[width=\columnwidth]{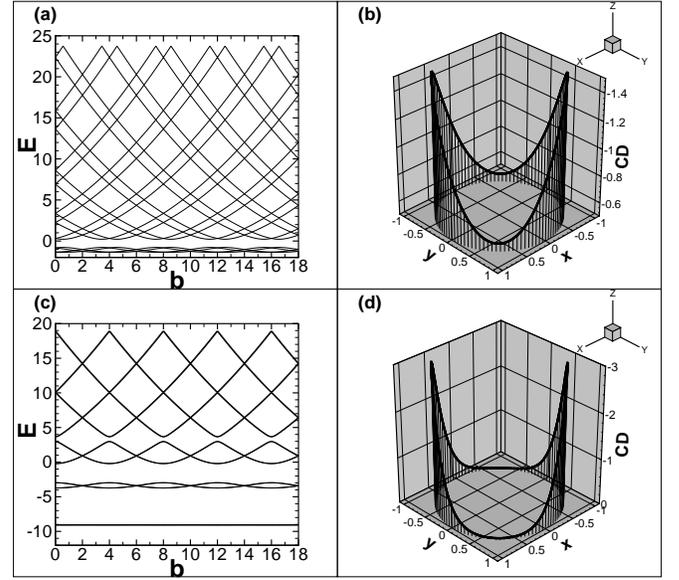}\caption{(a) Energy
spectrum of 1D ring while $\bar{\alpha} =2$, $\bar{\beta}=1$,
$\bar{g}=0$; (b) The charge density (CD) distribution of the lowest
single electron state in a 1D ring while $\bar{\alpha} =2$,
$\bar{\beta}=1$, $\bar{g}=0$, $b=0$; (c) Energy spectrum of 1D ring
while $\bar{\alpha} =4$, $\bar{\beta}=3$, $\bar {g}=0$; (d) The
charge density (CD) distribution of the lowest single electron state
in a 1D ring while $\bar{\alpha} =4$, $\bar{\beta}=3$, $\bar{g}=0$,
$b=0$. The charge density is in units of $e/2\pi a$.}
\label{fig:alphaneqbeta}
\end{figure}

Generally, the symmetry of the Hamiltonian shown in the previous
subsection no longer exists when $\left| \bar{\alpha}\right| \neq
\left| \bar{\beta}\right| $, even for $\bar{g}=0$. We show the
electron spectra for $\bar{\alpha}=2$, $\bar{\beta}=1$, $\bar{g}=0$
and $\bar{\alpha}=4$, $\bar{\beta}=3$, $\bar {g}=0$ in
Fig.~\ref{fig:alphaneqbeta}(a) and Fig.~\ref{fig:alphaneqbeta}(c),
respectively.  The energy gaps increase with increasing SOI
strengths. But the spin splitting in the two spectra is quite
different. It is interesting to note that the energy spectrum
becomes spin degenerate again when the two SOI strengths are tuned
to proper values even though they are different.
Figs.~\ref{fig:alphaneqbeta}(b) and \ref{fig:alphaneqbeta}(d) show
the distribution of charge density for different strength RSOI and
DSOI. The electron is localized along the ring due to the periodic
potential $\frac {\bar{\alpha}\bar{\beta}}{2}\sin2\varphi$. The
electron density distribution becomes more localized with increased
potential height, i.e., the product of the strengths of RSOI and
DSOI (see Eq.~(\ref{1dhami})).

\begin{figure}[ptb]
\includegraphics[width=\columnwidth]{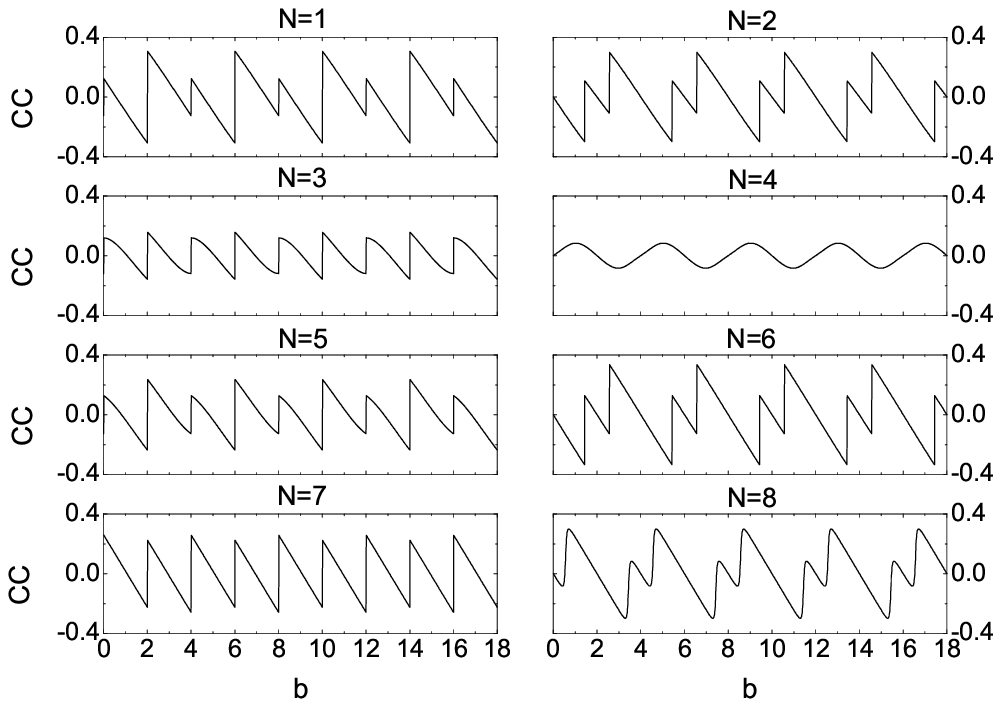}\caption{The persistent CC in a
1D ring with different numbers of electrons $N$ \textit{vs} magnetic
field $b$ while $\bar{\alpha}=2$, $\bar{\beta}=1$, $\bar{g}=0$. The
persistent CC is in units of \ $2NE_{0}/\Phi_{0}$.} \label{fig:cc21}
\end{figure}

\begin{figure}[ptb]
\includegraphics[width=\columnwidth]{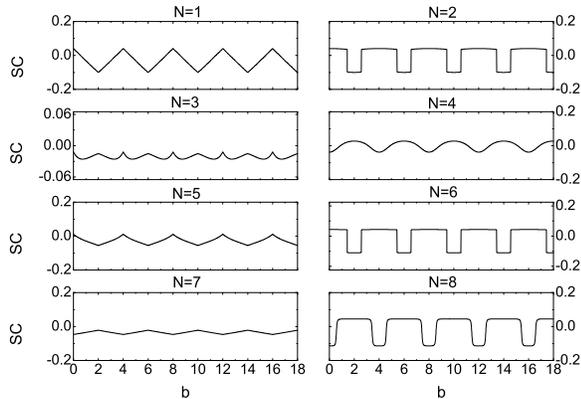}\caption{Same as
Fig.~\ref{fig:cc21}, but for the persistent SC. The persistent SC is
in units of $NE_{0}/2\pi$.}\label{fig:sc21}
\end{figure}

The persistent CC and SC are plotted in Fig.~\ref{fig:cc21} and
Fig.~\ref{fig:sc21}, respectively. We find that the oscillations of
the persistent CC and SC become smooth and weak due to the gaps in
the energy spectrum, especially when the Fermi level locates near
the largest energy gap ($N=4$). The persistent current (CC or SC)
oscillation no longer consists of linear segments since the
parabolic behavior of the energy levels disappears due to the
periodic potential $\frac{\bar{\alpha}\bar{\beta }}{2}\sin2\varphi$
in the 1D Hamiltonian (see Eq.~(\ref{1dhami})).

\begin{figure}[ptb]
\includegraphics[width=\columnwidth]{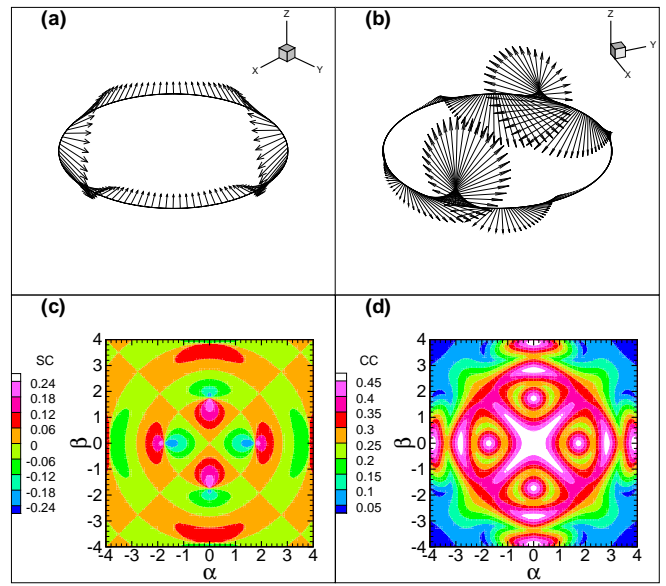}\caption{(color online) (a) Local spin
orientation $S(\bm{r})$ for the lowest spin-up level while
$\bar{\alpha}=2$, $\bar{\beta}=1$, $\bar{g}=0$, $b=2$; (b) Local
spin orientation $S(\bm{r})$ for the lowest spin-up level while
$\bar{\alpha}=4$, $\bar{\beta}=3$, $\bar {g}=0$, $b=2$; (c) The
persistent SC in a 1D ring with different RSOI and DSOI strengths
when the magnetic field $b$ is 2; (d) The oscillation amplitude of
the persistent CC in a 1D ring with different RSOI and DSOI
strengths. In Fig.~\ref{fig:contour}(c) and
Fig.~\ref{fig:contour}(d) we set $\bar{g}=0$ and $N=8$. The
persistent CC (SC) is in units of $2NE_{0}/\Phi_{0}$
($NE_{0}/2\pi$).}\label{fig:contour}
\end{figure}

The local spin orientation $S(\bm{r})$ also reveals the interplay
between the RSOI and DSOI since $S(\bm{r})^{R}$ is quite different
from $S(\bm{r})^{D}$. According to Eq.~(\ref{1dhami}), the interplay
between the RSOI and DSOI is divided into two parts, i.e.,
$\frac{\bar{\alpha}}{2}\sigma_{r}-\frac
{\bar{\beta}}{2}\sigma_{\varphi}( -\varphi)$ in the kinetic term and
the periodic potential $\frac{\bar{\alpha}\bar{\beta}
}{2}\sin2\varphi$. The first part makes the direction of the local
spin orientation vary along the ring, and the second part leads to
the electron localization (see Figs.~\ref{fig:contour}(a) and
\ref{fig:contour}(b)). When the SOI strengths increase, the spin
orientation exhibits rapid variation due to the enhancement of the
interplay between the RSOI and DSOI.

In Fig.~\ref{fig:contour}(c), we plot the persistent SC as a
function of RSOI strength $\bar{\alpha}$ and DSOI strength
$\bar{\beta}$ for a fixed number of electrons $N=8$ and magnetic
field $b=2$. The contour plot shows interesting symmetry. It is
symmetric (antisymmetric) with respect to the lines $\bar{\alpha}=0$
and $\bar{\beta}=0$ ($\bar{\alpha}=\pm \bar{\beta}$). From this
figure we find that the maxima and minima of the persistent SC occur
while only one of the two types of the SOI exists. That is because
the effects of the RSOI and DSOI on spin splitting tend to cancel
each other. The persistent SC becomes zero when the
strengths of the RSOI and DSOI are equal to each other ($\bar{\alpha}=\pm \bar{\beta}%
$). This corresponds to the spin degenerate case discussed before.
Besides the two orthogonal lines ($\bar{\alpha}=\pm \bar{\beta}$) on
the $\bar{\alpha }$-$\bar{\beta}$ plane there are many circle-like
closed curves on which the persistent SC disappears. These zero-SC
lines intersect the $\bar{\alpha}$ axis at those points $(\pm
\sqrt{m^{2}-1},0)$ and the $\bar{\beta}$ axis at $(0,\pm
\sqrt{m^{2}-1})$ where $m=1, 2, 3,\ldots$. The persistent SC
disappears because the energy spectrum becomes degenerate again when
the strengths of RSOI and DSOI are tuned to proper values even
though they are not equal. The contour plot of the oscillation
amplitude of the persistent CC for a fixed number of electrons $N=8$
is shown in Fig.~\ref{fig:contour}(d). The oscillation amplitude as
a function of $\bar{\alpha}$ and $\bar{\beta}$ is symmetric with
respect to the lines $\bar{\alpha}=0$ and $\bar{\beta}=0$ and
$\bar{\alpha }=\pm \bar{\beta}$. When only one type of SOI appears,
the maximum of the persistent CC oscillates with increased SOI
strength. When both types of SOI are included, the maximum of the
persistent CC decays since the interplay between the RSOI and DSOI
leads to a periodic potential along the ring which results in the
gaps in the energy spectrum, consequently smoothening and weakening
the oscillation of the persistent CC.

\subsection{Finite width effects}

\begin{figure}[ptb]
\includegraphics[width=\columnwidth]{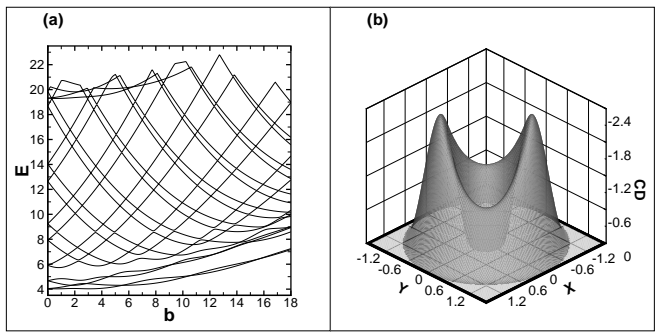}\caption{(a) Energy spectrum
for 2D hard wall ring with width $d=1.33$ while $\bar{\alpha}=2$,
$\bar{\beta }=1$, $\bar{g}=0$; (b) The charge density (CD)
distribution of the lowest single electron state in a 2D hard wall
ring with $d=1.33$ while $\bar{\alpha }=2$, $\bar{\beta}=1$,
$\bar{g}=0$, $b=0$. The charge density is in units of $e/2\pi
a^{2}$.}\label{fig:2dspec}
\end{figure}

Now we turn to consider a mesoscopic ring with finite width. When a
2D ring is thin enough, its characteristics are almost the same as
those of a 1D ring since the second radial levels are too high to be
occupied and the compressing effect of the magnetic field on the
radial wave function is negligible. Here we consider the wide ring
case.

The energy spectrum for a 2D ring with a finite width while
$\bar{\alpha}=2$, $\bar{\beta}=1$ and $\bar{g}=0$ is plotted in
Fig.~\ref{fig:2dspec}(a). The second radial mode can be seen in the
top of the energy spectrum. The compressing effect of the magnetic
field on the radial wave function accounts for the increase of the
energy with increasing magnetic fields. Like the 1D case, the
interplay between the RSOI and DSOI leads to an effective periodic
potential (see Eq.~(\ref{2dh})), resulting in energy gaps that
depend on the magnetic field. The charge density distribution of a
single electron in a 2D ring at $b=0$ is shown in
Fig.~\ref{fig:2dspec}(b). We find that the electron probability is
localized due to the $\sin2\varphi$ potential arising from the
interplay between the RSOI and DSOI.

\begin{figure}[ptb]
\includegraphics[width=\columnwidth]{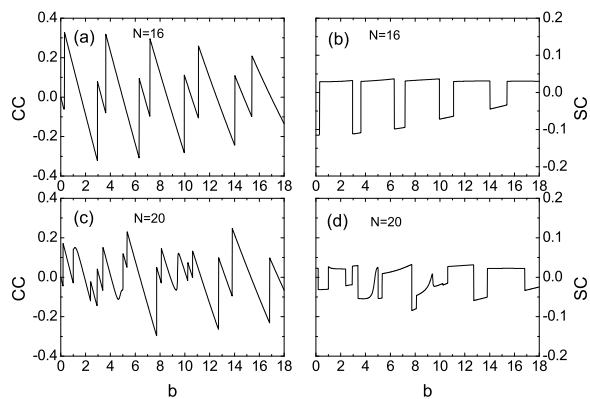}\caption{(a) The
perisistent CC in a 2D ring \textit{vs} magnetic field $b$ while
$\bar{\alpha }=2$, $\bar{\beta}=1$, $\bar{g}=0$, the width $d=1.33$
and $N=16$; (b) Same as (a) but for the persistent SC; (c) Same as
(a) except $N=20$; (d) Same as (b) except $N=20$. The persistent CC
(SC) is in units of
$2NE_{0}/\Phi_{0}$ ($NE_{0}/2\pi$).}%
\label{fig:2dcurrent}%
\end{figure}

The persistent CC and SC in the 2D wide ring as a function of
magnetic field are plotted in Fig.~\ref{fig:2dcurrent}. The number
of the electrons is tuned to detect the effect of the second radial
mode. When we consider the contribution of the lowest 16 levels, the
oscillations of the persistent CC and SC have a profile similar to
that in the 1D ring because the second radial mode has not been
involved yet. But when we increase the number of the electrons to
20, the quasi-periodicity of the persistent CC and SC as functions
of the magnetic field $b$ is destroyed. This feature arises from the
contribution of the second radial mode.

\section{\label{sec:summary}SUMMARY}

We have conducted a theoretical investigation of the spin states and
persistent CC and SC in mesoscopic rings with spin-orbit
interactions. We have demonstrated theoretically that the
Hamiltonian of the RSOI alone is mathematically equivalent to that
of the DSOI alone by a unitary transformation $T$. This property
results in the degenerate energy spectrum for equal strength RSOI
and DSOI. The interplay between the RSOI and DSOI leads to an
effective periodic potential $\frac
{\bar{\alpha}\bar{\beta}}{2}\sin2\varphi$. This periodic potential
results in gaps in the energy spectrum, and smoothens and weakens
the oscillations of the persistent CC and SC. The charge density and
the local spin orientation $S(\bm{r})$ are localized along the ring
due to the effect of the periodic potential. Higher radical modes
become involved as the ring width increases, destroying the
periodicity of the persistent CC and SC oscillations.

\begin{acknowledgments}
This work was supported by the NSFC Grant No. 60376016, 60525405.
\end{acknowledgments}

\appendix*

\section{The Hamiltonian and available analytical solutions}

The dimensionless Hamiltonian for a 2D ring reads
\begin{equation}
H_{2D}=H_{k}+H_{R}+H_{D}+H_{Z}+V(r),
\end{equation}
where the kinetic term $H_{k}=(\bm{e}_{r}k_{r}+\bm{e}_{\varphi}k_{\varphi
})^{2}$, the Rashba term $H_{R}=\bar{\alpha}(\sigma_{r}k_{\varphi}%
-\sigma_{\varphi}k_{r})$, the Dresselhaus term
$H_{D}=\bar{\beta}[\sigma
_{r}(-\varphi)k_{r}-\sigma_{\varphi}(-\varphi)k_{\varphi}]$, and the
Zeeman term $H_{Z}=\bar{g}b\sigma_{z}/2$. $V(r)$ is the radial
confining potential.
\begin{align}
H_{2D} &  =-\frac{\partial^{2}}{\partial r^{2}}-\frac{1}{r}\frac{\partial
}{\partial r}+V(r)+k_{\varphi}^{2}\nonumber \\
&  +\left[  \bar{\alpha}\sigma_{r}-\bar{\beta}\sigma_{\varphi}(-\varphi
)\right]  k_{\varphi}\nonumber \\
&  +\frac{i}{2r}\left[  -\bar{\alpha}\sigma_{\varphi}+\bar{\beta}\sigma
_{r}(-\varphi)\right]  +\frac{1}{2}\bar{g}b\sigma_{z}\nonumber \\
&  +\left(  k_{r}-\frac{i}{2r}\right)  \left[  -\bar{\alpha}\sigma_{\varphi
}+\bar{\beta}\sigma_{r}(-\varphi)\right]  \nonumber \\
&  =-\frac{\partial^{2}}{\partial r^{2}}-\frac{1}{r}\frac{\partial}{\partial
r}+V\left(  r\right)  \nonumber \\
&  +\left(  k_{r}-\frac{i}{2r}\right)  \left[  -\bar{\alpha}\sigma_{\varphi
}+\bar{\beta}\sigma_{r}(-\varphi)\right]  \nonumber \\
&  +\left[  k_{\varphi}+\frac{\bar{\alpha}}{2}\sigma_{r}-\frac{\bar{\beta}}%
{2}\sigma_{\varphi}\left(  -\varphi \right)  \right]  ^{2}\nonumber \\
&  -\left[  \frac{\bar{\alpha}}{2}\sigma_{r}-\frac{\bar{\beta}}{2}%
\sigma_{\varphi}(-\varphi)\right]  ^{2}+\frac{1}{2}\bar{g}b\sigma
_{z}\nonumber \\
&  =-\frac{\partial^{2}}{\partial r^{2}}-\frac{1}{r}\frac{\partial}{\partial
r}+V(r)\nonumber \\
&  +\left(  k_{r}-\frac{i}{2r}\right)  \left[  -\bar{\alpha}\sigma_{\varphi
}+\bar{\beta}\sigma_{r}(-\varphi)\right]  \nonumber \\
&  +\left[  k_{\varphi}+\frac{\bar{\alpha}}{2}\sigma_{r}-\frac{\bar{\beta}}%
{2}\sigma_{\varphi}(-\varphi)\right]  ^{2}\nonumber \\
&  -\frac{\bar{\alpha}^{2}+\bar{\beta}^{2}}{4}+\frac{\bar{\alpha}\bar{\beta}%
}{2}\sin2\varphi+\frac{1}{2}\bar{g}b\sigma_{z} \label{2dh}.
\end{align}
Specifically we write $H_{2D}=H_{0}+H_{1}$, where $H_{0}=-\frac{\partial^{2}%
}{\partial r^{2}}-\frac{1}{r}\frac{\partial}{\partial r}+V(r)$. The
correct 1D Hamiltonian $H$ can be obtained by evaluating the
expectation of $H_{1}$ in the lowest radial mode of
$H_{0}$.~\cite{Meijer-2002-033107} In the limit of a very narrow
ring, we can set $r$ to be a constant value ($r=1$) and the
following equation will hold for an arbitrarily given confining
potential $V(r)$:
\begin{equation}
\left \langle \rho_{0}\right \vert \frac{\partial}{\partial r}+\frac{1}%
{2r}\left \vert \rho_{0}\right \rangle =0.
\end{equation}
Here $\rho_{0}$ is the lowest radial mode for $V(r)$. Now we can
write the 1D Hamiltonian explicitly. We get
\begin{align}
H &  =\left[  -i\frac{\partial}{\partial \varphi}+\frac{b}{4}+\frac{\bar
{\alpha}}{2}\sigma_{r}-\frac{\bar{\beta}}{2}\sigma_{\varphi}(-\varphi)\right]
^{2}\nonumber \\
&  -\frac{\bar{\alpha}^{2}+\bar{\beta}^{2}}{4}+\frac{\bar{\alpha}\bar{\beta}%
}{2}\sin2\varphi+\frac{1}{2}\bar{g}b\sigma_{z}.\label{a1dhami}%
\end{align}

A unitary operator $T= \begin{bmatrix}
0 & \exp[-i\pi/4]\\
-\exp[i\pi/4] & 0
\end{bmatrix}
$ is defined, and we have $T^{\dag}=T^{-1}=-T$. By applying this
unitary operator, the Hamiltonian becomes
\begin{align}
THT^{\dag}  &  =\left[  -i\frac{\partial}{\partial \varphi}+\frac{b}{4}%
+\frac{\bar{\beta}}{2}\sigma_{r}-\frac{\bar{\alpha}}{2}\sigma_{\varphi}(
-\varphi)\right] ^{2}\nonumber \\
&  -\frac{\bar{\alpha}^{2}+\bar{\beta}^{2}}{4}+\frac{\bar{\alpha}\bar{\beta}
}{2}\sin2\varphi-\frac{1}{2}\bar{g}b\sigma_{z}\text{.}%
\end{align}
Thus the Hamiltonian in which $\bar{\alpha}=a$, $\bar{\beta }=b$,
$\bar{g}=c$ is mathematically equivalent to that in which
$\bar{\alpha}=b$, $\bar {\beta}=a$, $\bar{g}=-c$.

The Hamiltonian in matrix form is
\begin{align}
H  &  =
\begin{bmatrix} H_{11} & H_{12}\\ H_{21} & H_{22} \end{bmatrix} \text{,
where}\nonumber \\
H_{11}  &  =\left(  -i\frac{\partial}{\partial \varphi}+\frac{b}{4}\right)
^{2}+\bar{g}b/2 ,\nonumber \\
H_{12}  &  =\bar{\alpha}e^{-i\varphi}\left( -i\frac{\partial}{\partial \varphi
}+\frac{b}{4}-\frac{1}{2}\right)  +i\bar{\beta}e^{i\varphi}\left(
-i\frac{\partial}{\partial \varphi}+\frac{b}{4}+\frac{1}{2}\right)
,\nonumber \\
H_{21}  &  =\bar{\alpha}e^{i\varphi}\left(  -i\frac{\partial}{\partial \varphi
}+\frac{b}{4}+\frac{1}{2}\right)  -i\bar{\beta}e^{-i\varphi}\left(
-i\frac{\partial}{\partial \varphi}+\frac{b}{4}-\frac{1}{2}\right)
,\nonumber \\
H_{22}  &  =\left(  -i\frac{\partial}{\partial \varphi}+\frac{b}{4}\right)
^{2}-\bar{g}b/2 .
\end{align}
To solve the equation $H\Psi=E\Psi$, we expand the wavefunction $\Psi$ as
$\Psi= \begin{pmatrix}
\Psi_{1}\\
\Psi_{2}
\end{pmatrix}
=\sum_{m} \begin{pmatrix}
a_{m}\\
b_{m}
\end{pmatrix}
\Theta_{m}(\varphi)$, where $\Theta_{m}(\varphi)=\frac{1}{\sqrt{2\pi}}%
\exp \left[  im\varphi \right] $. The secular equation becomes
\begin{equation}
\left \{
\begin{array}
[c]{c}%
b_{m+1}\bar{\alpha}(m+\frac{b}{4}+\frac{1}{2})+ib_{m-1} \bar{\beta}(
m+\frac{b}{4}-\frac{1}{2})\\
=\left[ E-( m+\frac{b}{4})^{2}-\bar{g}b/2\right] a_{m}\\
a_{m-1}\bar{\alpha}(m+\frac{b}{4}-\frac{1}{2})-ia_{m+1} \bar{\beta}(m+\frac
{b}{4}+\frac{1}{2})\\
=\left[ E-( m+\frac{b}{4})^{2}+\bar{g}b/2\right] b_{m}%
\end{array}
\right. .\label{secular}%
\end{equation}
Generally, we can write the Hamiltonian in an infinite quintuple
diagonal matrix form based on Eq.~(\ref{secular}).
\begin{widetext}
\begin{align}
& \begin{bmatrix}
\ddots & \ddots & 0 & \ddots & 0 & 0 & 0 & 0\\
\ddots & \left( \frac{b}{4}-1\right)^{2}+\frac{\bar{g}b}{2} & 0 & 0
& \bar{\alpha
}\left(  \frac{b}{4}-\frac{1}{2}\right)   & 0 & 0 & 0\\
0 & 0 & \left(  \frac{b}{4}-1\right)  ^{2}-\frac{\bar{g}b}{2} &
-i\bar{\beta}\left(
\frac{b}{4}-\frac{1}{2}\right)   & 0 & 0 & 0 & 0\\
\ddots & 0 & i\bar{\beta}\left( \frac{b}{4}-\frac{1}{2}\right)   &
\left(\frac{b}{4}\right)^{2}+\frac{\bar
{g}b}{2} & 0 & 0 & \bar{\alpha}\left( \frac{b}{4}+\frac{1}{2}\right)   & 0\\
0 & \bar{\alpha}\left( \frac{b}{4}-\frac{1}{2}\right)   & 0 & 0 &
\left(\frac{b}{4}\right)^{2}-\frac{\bar
{g}b}{2} & -i\bar{\beta}\left(  \frac{b}{4}+\frac{1}{2}\right)   & 0 & \ddots \\
0 & 0 & 0 & 0 & i\bar{\beta}\left(  \frac{b}{4}+\frac{1}{2}\right) &
\left(
\frac{b}{4}+1\right)  ^{2}+ \frac{\bar{g}b}{2} & 0 & 0\\
0 & 0 & 0 & \bar{\alpha}\left(  \frac{b}{4}+\frac{1}{2}\right)   & 0
& 0 & \left(
\frac{b}{4}+1\right)  ^{2}-\frac{\bar{g}b}{2} & \ddots \\
0 & 0 & 0 & 0 & \ddots & 0 & \ddots & \ddots
\end{bmatrix} \nonumber \\
& \begin{bmatrix}
\vdots \\
a_{-1}\\
b_{-1}\\
a_{0}\\
b_{0}\\
a_{1}\\
b_{1}\\
\vdots
\end{bmatrix}
=E
\begin{bmatrix}
\vdots \\
a_{-1}\\
b_{-1}\\
a_{0}\\
b_{0}\\
a_{1}\\
b_{1}\\
\vdots
\end{bmatrix}.\label{quintuple}
\end{align}
\end{widetext}
We consider three different cases. For the first and second cases,
in which the quintuple diagonal matrices are reducible, analytical
solutions can be obtained.

a) $\bar{\alpha}\neq0$, $\bar{\beta}=0$;
\begin{align}
&
\begin{bmatrix} (m+b/4)^{2}+\bar{g}b/2 & \bar{\alpha}(m+b/4+1/2) \\ \bar{\alpha}( m+b/4+1/2) & (m+b/4+1)^{2}-\bar{g}b/2 \end{bmatrix} \begin{bmatrix} a_{m}\\ b_{m+1} \end{bmatrix}\nonumber \\
& =E \begin{bmatrix} a_{m}\\ b_{m+1} \end{bmatrix}.
\end{align}
The eigenvalues are
\begin{equation}
\label{rspectrum}E_{n,\sigma}^{R}=\left(
n+\frac{b}{4}+\frac{\sigma}{2} -\frac{\sigma}{2\cos
\theta_{n,\sigma}}\right)  ^{2}-\frac{\tan^{2}\theta_{n,\sigma}}
{4}+\sigma \frac{\bar{g}b}{2\cos \theta_{n,\sigma}},
\end{equation}
where $\tan \theta_{n,\sigma}=\frac{\bar{\alpha}(n+b/4+\sigma/2)
}{n+b/4+\sigma/2-\bar{g}b/2}$. The corresponding eigenvectors are
\begin{equation}
\label{rups}\Psi_{n,\uparrow}^{R}=\frac{1}{\sqrt{2\pi}}e^{i(n+1/2)
\varphi}
\begin{pmatrix} \cos \frac{\theta_{n,\uparrow}}{2}e^{-i\varphi/2}\\ -\sin \frac{\theta_{n,\uparrow}}{2}e^{i\varphi/2} \end{pmatrix}
\end{equation}
and
\begin{equation}
\label{rdowns}\Psi_{n,\downarrow}^{R}=\frac{1}{\sqrt{2\pi}}e^{i(n-1/2)\varphi}
\begin{pmatrix} \sin \frac{\theta_{n,\downarrow}}{2}e^{-i\varphi/2}\\ \cos \frac{\theta_{n,\downarrow}}{2}e^{i\varphi/2} \end{pmatrix}.
\end{equation}
The local spin orientations for eigenstates are
\begin{align}
&  S(\bm{r})_{n,\uparrow}^{R}\nonumber \\
&  =\frac{\hbar}{4\pi a}\left[  \sin(-\theta_{n,\uparrow})( \cos \varphi \mathbf{e}%
_{x}+\sin \varphi \mathbf{e}_{y}) +\cos(
-\theta_{n,\uparrow})\mathbf{e}_{z}\right]
\end{align}
and
\begin{align}
& S(\bm{r})_{n,\downarrow}^{R}\nonumber \\
& =\frac{\hbar}{4\pi a}\left[  \sin(\pi-\theta_{n,\downarrow})( \cos \varphi \mathbf{e}%
_{x}+\sin \varphi \mathbf{e}_{y}) +\cos(
\pi-\theta_{n,\downarrow})\mathbf{e}_{z}\right] .
\end{align}

b) $\bar{\alpha}=0$, $\bar{\beta}\neq0$;
\begin{align}
&
\begin{bmatrix} (m+b/4+1)^{2}+\bar{g}b/2 & i\bar{\beta}(m+b/4+1/2)\\ -i\bar{\beta}(m+b/4+1/2) & (m+b/4)^{2}-\bar{g}b/2 \end{bmatrix} \begin{bmatrix} a_{m+1}\\ b_{m} \end{bmatrix}\nonumber \\
&  =E \begin{bmatrix} a_{m+1}\\ b_{m} \end{bmatrix}.
\end{align}
The eigenvalues are
\begin{equation}
\label{dspectrum}E_{n,\sigma}^{D}=\left(
n+\frac{b}{4}-\frac{\sigma}{2}
+\frac{\sigma}{2\cos \eta_{n,\sigma}}\right) ^{2}-\frac{\tan^{2}\eta_{n,\sigma}}{4}%
+\sigma \frac{\bar{g}b}{2\cos \eta_{n,\sigma}},
\end{equation}
where $\tan \eta_{n,\sigma}=\frac{\bar{\beta}(n+b/4-\sigma/2)
}{n+b/4-\sigma/2+\bar{g}b/2}$. The corresponding eigenvectors are
\begin{equation}
\label{dups}\Psi_{n,\uparrow}^{D}=\frac{1}{\sqrt{2\pi}}e^{i(n-1/2)\varphi}
\begin{pmatrix} \cos \frac{\eta_{n,\uparrow}}{2}e^{i\varphi/2}\\ -i\sin \frac{\eta_{n,\uparrow}}{2}e^{-i\varphi/2} \end{pmatrix}
\end{equation}
and
\begin{equation}
\label{ddowns}\Psi_{n,\downarrow}^{D}=\frac{1}{\sqrt{2\pi}}e^{i(n+1/2)
\varphi}
\begin{pmatrix} -i\sin \frac{\eta_{n,\downarrow}}{2}e^{i\varphi/2}\\ \cos \frac{\eta_{n,\downarrow}}{2}e^{-i\varphi/2} \end{pmatrix}.
\end{equation}
The local spin orientations for the eigenstates are
\begin{align}
&  S(\bm{r})_{n,\uparrow}^{D}\nonumber \\
&  =\frac{\hbar}{4\pi a}\left[  \sin(-\eta_{n,\uparrow})( \sin \varphi \mathbf{e}%
_{x}+\cos \varphi \mathbf{e}_{y})+\cos(
-\eta_{n,\uparrow})\mathbf{e}_{z}\right]
\end{align}
and
\begin{align}
&  S(\bm{r})_{n,\downarrow}^{D}\nonumber \\
&  =\frac{\hbar}{4\pi a}\left[  \sin( \pi-\eta_{n,\downarrow})( \sin \varphi \mathbf{e}%
_{x}+\cos \varphi \mathbf{e}_{y}) +\cos(
\pi-\eta_{n,\downarrow})\mathbf{e}_{z}\right] .
\end{align}

c) $\bar{\alpha}\neq0$, $\bar{\beta}\neq0$.

While both RSOI and DSOI have nonvanishing strengths, we cannot
reduce the infinite quintuple diagonal matrix shown in
Eq.~(\ref{quintuple}) into a more compact form. Thus analytical
solutions do not exist. We give numerical results instead.
\bibliography{myref}

\end{document}